\documentclass[aps,prl,showpacs,twocolumn,superscriptaddress]{revtex4}

\usepackage{graphicx}
\usepackage{longtable}
\usepackage[ansinew]{inputenc}

\begin{document}

  \bibliographystyle{prsty}

  \title{Controlled Contact to a C$_{60}$ Molecule}
  \author{N. N\'{e}el}
  \author{J. Kr\"{o}ger}\email{kroeger@physik.uni-kiel.de}
  \author{L. Limot}
  \affiliation{Institut für Experimentelle und Angewandte Physik, Christian-Albrechts-Universität zu Kiel, D-24098 Kiel, Germany}
  \author{T. Frederiksen}
  \author{M. Brandbyge}
  \affiliation{MIC -- Department of Micro and Nanotechnology, NanoDTU, Technical University of Denmark, DK-2800 Kongens Lyngby, Denmark}
  \author{R. Berndt}
  \affiliation{Institut für Experimentelle und Angewandte Physik, Christian-Albrechts-Universität zu Kiel, D-24098 Kiel, Germany}
  \date{\today}

  \begin{abstract}
    The conductance of $\text{C}_{60}$ on Cu(100) is investigated with a
    low-temperature scanning tunneling microscope. At the transition from
    tunneling to the contact regime the conductance of $\text{C}_{60}$ adsorbed
    with a pentagon-hexagon bond rises rapidly to $\approx 0.25$ conductance
    quanta $\text{G}_0$. An abrupt conductance jump to $\text{G}_0$ is observed
    upon further decreasing the distance between the instrument's tip and the
    surface. {\it Ab-initio} calculations within density functional theory
    and non-equilibrium Green's function techniques explain the experimental
    data in terms of the conductance of an essentially undeformed
    $\text{C}_{60}$. From a detailed analysis of the crossover from tunneling
    to contact we conclude that the conductance in this region is strongly
    affected by structural fluctuations which modulate the tip-molecule
    distance.
  \end{abstract}


  \maketitle

  The mechanical and electronic properties of materials at the atomic scale
  are important in various areas of research which range from fundamentals of
  adhesion and friction to photosynthesis and signal transduction in molecular
  structures. Electronic transport through nanostructures may find applications
  in devices and is being investigated for semiconducting \cite{bjw_88} and
  metallic \cite{cjm_92,esc_97} constrictions, carbon nanotubes \cite{sfr_98},
  DNA \cite{mri_97,hwf_99,dpo_00,ayk_01}, and single metal atoms \cite{lli_05}.

  Scanning tunneling microscopy (STM) appears to be an ideal tool to study
  single molecule conductance in more detail. The structure under investigation
  -- a molecule along with its substrate -- can be imaged with sub-molecular
  precision prior to and after taking conductance data. Parameters such as
  molecular orientation or binding site, which are expected to significantly
  affect conductance properties, can thus be monitored. Moreover, specific
  parts of a molecule may be addressed to probe their role in electron
  transport, signal transduction or energy conversion. Another advantage of
  STM is the possibility to characterize to some extent the status of the
  second electrode, the microscope tip, by recording conductance data on
  clean metal areas. Consequently, STM can complement techniques like the
  mechanical break junction measurements.

  Scanning probe techniques have indeed been used to form point contacts
  between the tip and a metal surface whose quantized conductance was then
  investigated during forming and stretching of the contact \cite{jkg_87,jip_93,lol_94}.
  Taking advantage of the imaging capability of STM a recent experiment on
  single-adatom contacts \cite{lli_05} showed that tip-adatom contacts can be
  formed reproducibly without structural changes of tip or sample. Somewhat
  surprisingly, STM data for molecular point contacts are scarce.
  Joachim {\it et al.} \cite{cjo_95} used STM at ambient temperature and
  modeling to investigate the electrical contact to a $\text{C}_{60}$ molecule
  on Au(110). Sub-molecular features were not resolved in STM images of this
  pioneering work, likely owing to molecular rotation. The interaction between
  the tip and Cu-tetra-3,5 di-terbutyl-phenyl porphyrin adsorbed on Cu(211)
  was also investigated \cite{fmo_01}.

  Here we report on detailed measurements of the conductance $G$ of
  $\text{C}_{60}$ molecules adsorbed on Cu(100) in a low-temperature scanning
  tunneling microscope. By recording STM images we determined the molecular
  orientation and position and confirmed that it remained unchanged prior to
  and after these measurements despite the large currents (up to $30\,\mu\text{A}$)
  passed through a molecule. We also monitored the tip status by spectroscopy
  of nearby pristine Cu surface areas. The transition from the tunneling to
  the contact regime is signalled by a rapid rise of the conductance to
  $G\approx 0.25\,\text{G}_0$, where $\text{G}_0 = 2\,\text{e}^2/\text{h}$.
  When approaching the tip further towards the molecule a jump up to
  $G\approx 1\,\text{G}_0$ is observed. The experimental data are modeled
  with density functional theory (DFT) and non-equilibrium Green's function
  (NEGF) techniques. Our theory captures the important characteristics of the
  experiment and explains the underlying physics. Mechanical contact between
  tip and molecule is found to result in a conductance lower than $\text{G}_0$
  in agreement with the experimental observation. The forces on the tip and
  the molecule at close proximity lead to small changes of the atomic positions
  without affecting significantly the spherical shape of the $\text{C}_{60}$
  molecule. Further, the behavior in the crossover region from tunneling to
  contact is understood in terms of a fluctuation between two microscopic
  configurations, one with and the other without a chemical tip-molecule
  bond. These experimental and theoretical findings differ from those reported
  for $\text{C}_{60}$ on Au(110) \cite{cjo_95}.

\begin{figure}[t]
    \includegraphics[bbllx=0,bblly=0,bburx=453,bbury=453,width=50mm,clip=]{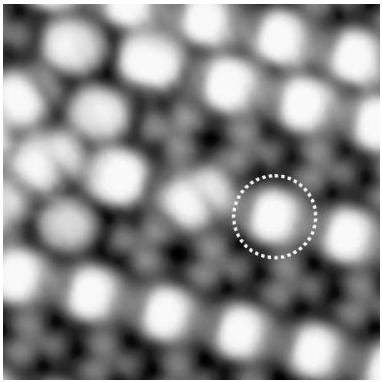}
    \caption[fig1]{(color online) Constant-current STM image of
    Cu(100)-$\text{C}_{60}$ at $8\,\text{K}$. (Sample voltage
    $V=1.7\,\text{V}$, tunneling current $I=1\,\text{nA}$, scan size
    $40\,\text{\AA}\times 40\,\text{\AA}$). A dashed circle indicates the
    $\text{C}_{60}$ orientation on which we performed the conductance
    measurements.}
    \label{fig1}
  \end{figure}

\begin{figure}[t]
    \includegraphics[bbllx=0,bblly=0,bburx=770,bbury=500,width=85mm,clip=]{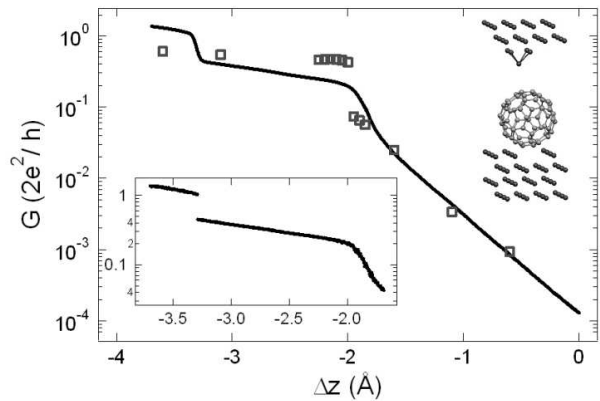}
    \caption[fig2]{(color online) Conductance $G$ in units of the quantum of
    conductance $\text{G}_0 = 2\,\text{e}^2/\text{h}$ plotted versus tip
    displacement $\Delta z$. Zero displacement corresponds to the tip position
    before freezing the feedback loop at $V=300\,\text{mV}$ and $I=3\,\text{nA}$.
    Experimental data are depicted as dots; calculated data are presented as
    squares. Upper inset: Theoretical setup for calculations. Lower inset:
    Single conductance curve revealing a conductance discontinuity at
    $\Delta z\approx -3.3\,\text{\AA}$.}
    \label{fig2}
  \end{figure}

  Our experiments were performed with a scanning tunneling microscope operated
  at $8\,\text{K}$ and in ultra-high vacuum with a base pressure of
  $10^{-9}\,\text{Pa}$. The Cu(100) surface as well as chemically etched
  tungsten tips were cleaned by annealing and argon ion bombardment.
  $\text{C}_{60}$ was evaporated from a tantalum crucible while keeping the
  residual gas pressure below $5\times 10^{-8}\,\text{Pa}$. An ordered
  $\text{C}_{60}$ superstructure was obtained by deposition onto the clean
  surface at room temperature and subsequent annealing to $500\,\text{K}$.
  Deposition rates were calibrated with a quartz microbalance to be
  $\approx 1\,\text{ML\,min}^{-1}$. We define a monolayer (ML) as one
  $\text{C}_{60}$ molecule per sixteen copper atoms. Tips were controllably
  indented into the substrate material prior to conductance measurements.
  Consequently, tips were most likely covered with copper. Only tips which
  exhibited submolecular resolution (Fig.\,\ref{fig1}) in constant-current
  STM images were used. We experienced that the conductance curves depend on
  the tip shape. While tips with presumably a single atom at the apex lead
  to data as presented in Fig.\,\ref{fig2}, blunt tips exhibit larger contact
  conductances.
  We made sure that in spite of the unusually high currents no significant
  voltage drop at the input impedance of the current-to-voltage converter
  occured. Thus, the decrease of the bias voltage at the tip-molecule
  junction was negligible.

  A typical constant-current STM image of annealed Cu(100)-$\text{C}_{60}$
  is shown in Fig.~\ref{fig1}. The molecules are arranged in a hexagonal
  lattice and exhibit a superstructure of bright and dim rows which is
  associated with a missing-row reconstruction of the copper surface \cite{mab_03}.
  Bright rows correspond to $\text{C}_{60}$ molecules in a missing Cu row
  while dim rows correspond to molecules located at double missing rows.
  Fig.~\ref{fig1} exhibits, similar to the case of $\text{C}_{60}$ on
  Ag(100) \cite{xlu_03}, four molecular orientations on Cu(100) \cite{nne_06}.

  To study theoretically the Cu(100)-$\text{C}_{60}$ system in the presence
  of an STM tip we use the \textsc{Siesta} \cite{jso_02} and \textsc{Transiesta}
  \cite{mbr_02} DFT packages \footnote{Electronic structure calculations
  are based on the generalized gradient approximation for the exchange-correlation
  functional, a single-$\zeta$ plus polarization basis for the valence
  electrons, a $200\,\text{Ry}$ cutoff energy for the real space grid integrations,
  and the $\Gamma$-point approximation. Core electrons are described with
  pseudopotentials. The conductance is calculated from the zero-bias
  transmission at the Fermi energy sampled over $3\times 3$ $\mathbf k$-points
  in the two-dimensional Brillouin zone in the transverse plane of the
  transport.}. The system is modeled by a periodic supercell
  containing one $\text{C}_{60}$ molecule on a $4\times 4$ representation of
  six Cu(100) layers with a single missing row surface. The tip represented
  by a Cu pyramid mounted on the reverse side of the surface film. We have
  illustrated this setup in the upper inset of Fig.~\ref{fig2}. To determine
  the microscopic arrangement at different tip-substrate distances we vary
  the length of the supercell in the direction perpendicular to the surface
  and relax both $\text{C}_{60}$ and tip atoms until all residual forces on
  these atoms are smaller than $0.02\,\text{eV}/\text{\AA}$. The conductance
  is finally determined from a calculation of the zero-bias transmission
  function of the junction by including DFT self-energies for the coupling to
  semi-infinite atomistic leads.

  In the following we discuss electron transport measurements through an
  individual $\text{C}_{60}$ which is adsorbed with its 5:6 bond, {\it i.\,e.},
  the molecule is oriented such as to exhibit a carbon-carbon bond between a
  carbon pentagon and a carbon hexagon at the top (see the molecule encircled
  by a dashed line in Fig.\,\ref{fig1}). Fig.~\ref{fig2} presents experimental
  (dots) and calculated (open squares) results for the conductance $G=I/V$
  (in units of $\text{G}_0$) on a logarithmic scale. Owing to the large number
  of data points ($\approx 1150$) dots overlap and appear as a line. The
  displacement axis shows the tip excursion towards the molecule with $\Delta z=0$
  corresponding to the position of the tip before opening the feedback loop of
  the instrument. The tip is then moved towards the molecule ($\Delta z<0$)
  by more than $3.5\,\text{\AA}$ while the current is simultaneously recorded
  to explore the evolution of the conductance of the tip-molecule junction
  in a wide range of distances between the tip and the molecule. Experimental
  data points represent averages over 500 measurements recorded at a sample
  voltage of $300\,\text{mV}$. Conductance curves recorded at voltages
  between $50\,\text{mV}$ and $600\,\text{mV}$ revealed a similar shape.

  Typical characteristics of the conductance curve are as follows. Between
  $\Delta z = 0$ and $\Delta z\approx -1.6\,\text{\AA}$ the conductance varies
  exponentially from $10^{-4}\,\text{G}_0$ to $\approx 0.025\,\text{G}_0$
  consistent with electron tunneling from tip to sample states. Starting from
  $\Delta z\approx -1.6\,\text{\AA}$ we observe deviations from the linear
  behavior in the semilog plot. A sharp increase of the conductance by a factor of 10 to
  $0.25\,\text{G}_0$ occurs within a displacement interval of
  $\approx 0.4\,\text{\AA}$. For comparison, in the tunneling regime this
  displacement leads to an increase of the conductance by only a factor of
  $3.5$. Further decrease of the tip-molecule distance augments the conductance
  although the slope is reduced by a factor of ten compared to the tunneling
  regime. At a displacement of $\Delta z\approx -3.3\,\text{\AA}$ a second
  rapid increase of the conductance to $\text{G}_0$ is observed. As evident
  from the lower inset of Fig.\,\ref{fig2} which displays a single conductance
  curve this rise is actually a discontinuity. Owing to small variability
  of the exact location of this jump, averaging over 500 instances leads to
  some broadening. Upon further approach, the conductance exhibits yet another
  very small increase with decreasing tip-molecule distance. We note that for
  tip excursions $\Delta z < -3.8\,\text{\AA}$, we usually observed instabilities
  and damage of the tip or sample.
  \begin{figure}[t]
    \includegraphics[bbllx=69,bblly=59,bburx=551,bbury=773,width=70mm,clip=]{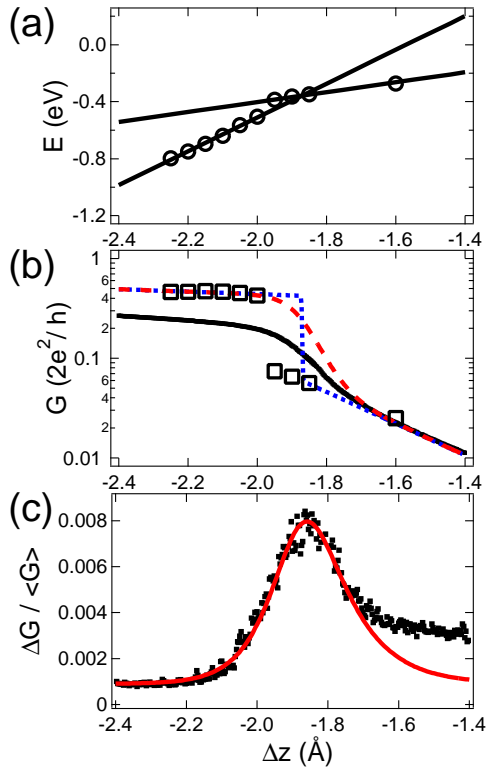}
    \caption[fig3]{(color online) (a) Calculated total energy differences
    versus tip displacement $\Delta z$ in the transition region from tunneling
    to contact. The data points (circles) fall on one of two straight lines
    corresponding to either a tunneling (smaller slope) or a contact (larger
    slope) configuration. (b) Experimental (dots) and theoretical (squares
    and dashed lines) conductance data in the tunneling-contact transition
    regime. Thin and thick dashed lines represent the theoretical conductance
    corresponding to a thermal average for a fluctuation between tunneling
    and contact configurations with $T=8\,\text{K}$ and $T=400\,\text{K}$,
    respectively (see text). (c) Ratio of the standard deviation $\Delta G$
    over the mean conductance $\langle G\rangle$ evaluated over $500$ conductance
    curves within the tunneling-contact transition regime. Full line: Calculated
    data for an effective temperature of $400\,\text{K}$ (data divided by
    200).}
    \label{fig3}
  \end{figure}

  The results of our calculations (squares in Fig.\,\ref{fig2}) describe most
  of the essential features of the experimental conductance data. The tunneling
  regime is reproduced with the experimentally measured slope. A rapid increase
  of the conductance occurs at $\Delta z\approx -2.0\,\text{\AA}$, leading to
  a conductance which is comparable to the experimental value and clearly
  lower than $\text{G}_0$ \footnote{The small quantitative difference between
  theory and experiment for the contact conductance might be related tip shape
  and tip position over the $\text{C}_{60}$ molecule.}. This rise of the
  conductance can be understood from the relaxed tip-molecule geometries.
  As the electrode separation is
  reduced by only $0.05\,\text{\AA}$, the tip-molecule distance shrinks from
  $3.18\,\text{\AA}$ to $2.34\,\text{\AA}$. This results in the formation of
  a chemical bond between the tip apex and the $\text{C}_{60}$ which hence
  effectively closes the tunneling gap. Concomitantly, the conductance increases
  by a factor of six. Around this instability point -- which defines the
  crossover from tunneling to contact -- we find that only small energy
  differences discriminate between the configurations with or without the
  tip-molecule bond. This is shown in Fig.\,\ref{fig3}a where the calculated
  zero-temperature data points are seen to fall on one of two straight lines
  that correspond to either a tunneling (smaller slope) or a contact (larger
  slope) configuration of the junction. At finite temperatures and under the
  non-equilibrium conditions imposed by the bias voltage, it is therefore
  likely that the junction will fluctuate between these different situations.
  From a couple of data points just before (after) the conductance jump we
  can extrapolate the distance dependence of the conductance $G_\mathrm{t}$
  ($G_\mathrm{c}$) and total energy $E_\mathrm{t}$ ($E_\mathrm{c}$)
  corresponding to a tunneling (contact) configuration. With these at hand
  we can establish the thermally averaged conductance over a fluctuation
  between these two situations according to
  \begin{equation}
    \overline{G}(z) = \frac{G_\mathrm{t}(z)e^{-\beta E_\mathrm{t}(z)}+G_\mathrm{c}(z)e^{-\beta E_\mathrm{c}(z)}}{e^{-\beta E_\mathrm{t}(z)}+e^{-\beta E_\mathrm{c}(z)}},
  \end{equation}
  where $\beta = 1/\text{k}_{\text{B}}T$ is the inverse temperature. The
  results of this procedure are shown in Fig.\,\ref{fig3}b with dashed lines
  corresponding to two different values for the effective temperature.
  With the temperature of the cryostat ($T=8\,\text{K}$) a sharp crossover
  from tunneling to contact is predicted to occur around
  $\Delta z = -1.87\,\text{\AA}$. The position of this jump agrees very well
  with that of the experimental data but its width is too narrow. However,
  if the effective temperature is increased to $T=400\,\text{K}$ the
  experimental width of the transition region is perfectly reproduced by our
  calculations. From an estimate of the maximal energy dissipation in the
  junction at the given bias voltage we find that this effective temperature
  is plausible. Further, the evaluated relative error of experimentally
  acquired conductances exhibits a maximum in the transition regime from
  tunneling to contact (see Fig.\,\ref{fig3}c) pointing at structural
  fluctuations which modulate the tip-molecule distance and thus the conductance.
  Except for absolute values this curve can be reproduced by our calculations.
  Additionally, the width of the transition depends on the bias voltage,
  {\it i.\,e.}, on the energy dissipation in the junction. These observations
  are strong indications that the fluctuation interpretation is correct.

  As a consequence, we arrive at a picture of the tip-$\text{C}_{60}$ contact
  which differs from the one reported by Joachim {\it et al.}\,\cite{cjo_95}
  for Au(110)-$\text{C}_{60}$.
  Some differences are highlighted below. Defining the conductance
  $G=1.3\times 10^{-4}\,\text{G}_0$, which is located in the tunneling regime,
  as a point of reference, we observe an exponential tunneling type of conductance
  variation over a range of $1.6\,\text{\AA}$ whereas mechanical contact
  along with an accelerated rise in conductance is reported for
  $\Delta z \approx -0.8\,\text{\AA}$ in Ref.\,\cite{cjo_95}. In our experiment
  on Cu(100), $G$ reaches $\text{G}_0$ at $\Delta z \approx -3.3\,\text{\AA}$
  whereas $G$ is still smaller than $\text{G}_0$ in Ref.\,\cite{cjo_95} for
  displacements as large as $-10\,\text{\AA}$ where the $\text{C}_{60}$ cage
  has already collapsed at $\Delta z \approx -5\,\text{\AA}$. Within our model,
  the deformation of the $\text{C}_{60}$ molecule in contact with the tip is
  small. The molecule remains almost spherical with only small relaxations of
  the carbon-carbon bond lengths (the diameter of the cage changes less than
  $4\,\%$).

  We finally comment on the experimentally observed discontinuous conductance
  jump to $1\,\text{G}_0$ at $\Delta z\approx -3.3\,\text{\AA}$. Since
  with blunt tips we experienced a continuous transition from tunneling to
  contact with a contact conductance of $\text{G}_0$, we attribute the
  discontinuous conductance change to a sudden rearrangement of the tip or
  molecule geometry leading to a higher number of conductance channels.

  In conclusion, we used low-temperature STM and theoretical modeling to
  investigate point contacts to $\text{C}_{60}$ on Cu(100). In the experiment,
  the junction is stable up to currents of $30\,\mu\text{A}$ and reproducible
  conductance data is obtained. When approaching the microscope's tip,
  deviations from tunneling are observed similar to those observed from single
  adatoms which are due to deformations of the tip. At contact, we find a
  conductance $G\approx 0.25\,\text{G}_0$.  Further decrease of the gap
  spacing leads to a discontinuous conductance change to $G=\text{G}_0$.
  From our modeling we infer that the controlled contact to a C$_{60}$
  molecule does not significantly deform its spherical shape. Moreover, we show
  that the conductance around the tip-molecule contact formation is strongly
  affected by a fluctuation between different microscopic configurations.

  N.~N., J.~K., L.~L., and R.~B.~thank C.\ Cepek (Laboratorio Nazionale TASC,
  Italy) for providing $\text{C}_{60}$ molecules. T.~F.~and M.~B.~thank the
  Danish Center for Scientific Computing (DCSC) for computational resources.

\end{document}